\documentclass[a4paper,11pt]{article}
\usepackage{jheppub}

\rightline{\small LMU-ASC 06/13}
\rightline{\small MPP-2013-22}

\title{Four K\"ahler Moduli Stabilisation in type IIB Orientifolds with
K3-fibred Calabi-Yau threefold compactification}

\author[a,b]{Dieter L\"ust}\author{and}
\author[a,c]{Xu Zhang}

\affiliation[a]{Max-Planck-Institute for Physics, F\"ohringer Ring
6, D-80805 Munich, Germany}
\affiliation[b]{Ludwig-Maximilians-Universit\"at, Arnold-Sommerfeld-Center,
Theresienstrasse 37, D-80333 Munich, Germany}
\affiliation[c]{State Key Laboratory of Theoretical Physics,
Institute of Theoretical Physics,
Chinese Academy of Sciences, Beijing 100190, China}

\emailAdd{dieter.luest@lmu.de}
\emailAdd{luest@mppmu.mpg.de}
\emailAdd{xuzhang@mpp.mpg.de}

\abstract{We present a concrete and consistent procedure to generate one kind
of non-perturbative superpotential, including the gaugino condensation
corrections and poly-instanton corrections, in type IIB orientifold
compactification with four K\"ahler Moduli. Then we use this kind of
superpotential
as well as the $\alpha^\prime$-corrections to K\"ahler potential to fix all of
the four K\"ahler moduli on a general Calabi-Yau manifold with typical K3-fibred
volume form. In our construction, the considered Calabi-Yau threefolds are
K3-fibred and admit at least one del Pezzo surface and one W-surface.
Searching through all existing four dimensional reflexive lattice polytopes, we
find 23 of them fulfilling all the requirements.}


\begin{document}

\maketitle

\section{Introduction}
The moduli parameters in string theory correspond to massless scalars in
4-dimensional
effective supergravity and hence will lead to long range interactions. The
couplings of these scalars to matter fields are in general not universal, which
implies that
different matter fields will obtain different accelerations from these long
range
forces. Obviously, this phenomenon violates the principle of equivalence, which
has been tested by the ratio of inertial to gravitational mass up to $10^{-13}$
\cite{0103036}. Therefore a ``fifth force'' must be very weak or sufficiently
short ranged, and a very natural consequence is that all of the moduli should be
massive. Furthermore, string theory loses any predictability, if the vacuum
expectation values of the moduli fields, especially for the volume modulus, can
take arbitrary values, since many physical parameters in the low energy  theory
depend
on the specific value of moduli.

In the type IIB  orientifold compactifications with $O7/O3$-planes,
there are two proposed mechanisms to stabilise all of the moduli, at least in
the case of a few K$\ddot{\textrm{a}}$hler moduli, i.e $h^{1,1}$ is small. One
is called KKLT strategy \cite{0301240}, and the other one is the LARGE volume
scenario(LVS) \cite{0502058}. In both cases, one first stabilises the
axion-dilaton and complex structure moduli by appropriate choice of background
fluxes, more concretely by the Gukov-Vafa-Witten superpotential induced by these
fluxes; then one fixes the values of the K$\ddot{\textrm{a}}$hler moduli by
non-perturbative effects such as D-brane instantons and gaugino condensation. 
Many explicit constructions in type IIB orientifold compactification have been
investigated for both cases, see for example
\cite{0609013,1105.2107,1208.1160,0808.0691}. The key difference between these
two mechanisms is that the LVS admits a non-supersymmetric anti-de Sitter
minimum
instead of the supersymmetric one in KKLT, and the fixed value of the Calabi-Yau
manifold volume is exponentially large with respect to the size of the ``small''
four-cycle, which is usually a del-Pezzo surface and supports the $D$-brane
instanton or the gaugino condensation. Furthermore the value of fixed ``small''
four-cycle volume is independent of the flux superpotential $W_0$ at fixed
$g_s$, which implies that this non-perturbative stabilisation of the
K$\ddot{\textrm{a}}$hler moduli will not disturb the complex structure
stabilisation. This also avoids the fine tuning of $W_0$ and the necessity
of a large gauge group as in the KKLT strategy. A more elaborated survey of
moduli
stabilisation mechanisms is presented in the appendix of \cite{0708.1873}.

The key requirement for a LVS model is to find a Calabi-Yau threefold with
$h^{1,2}>h^{1,1}>1$, where the volume can be expressed according to  the strong
`Swiss cheese' type or K3-fibred type Calabi-Yau threefolds. In addition, there
must be divisors which can support non-perturbative effects in the Calabi-Yau
threefold. In this case, it is possible to make some of the four-cycles small
while keeping the volume large. Even in the large volume limit, the induced
non-perturbative effects can compete against the $\alpha^\prime$-corrections to
the K$\ddot{\textrm{a}}$hler potential. Actually quite some progress has been
made in this context, especially for $h^{1,1}\le3$. For the case of strong
`Swiss cheese' type Calabi-Yau, one stabilises the volume plus one ``small''
four cycle by the $\alpha^\prime$-corrections and the non-perturbative effects,
and the third one is fixed by poly-instanton effects, e.g. \cite{1208.1160}. For
the case of a K3-fibred Calabi-Yau, the third one can be fixed either by the
string-loop effects to K\"ahler potential \cite{0808.0691}, or by the
poly-instanton effects \cite{1105.2107}.

Particularly in the case of a K3-fibred Calabi-Yau with poly-instanton
corrections, we may obtain an anisotropic extra dimensions and a TeV string
scale, which is advantageous for embedding extra-dimensional models of particle
physics into type IIB string theory, as shown in \cite{1105.2107}. And this case
is also very useful in the cosmology model constructions, for example the
single-field inflation models \cite{0808.0691,1110.6182}, the
double-field inflation models using the curvaton mechanism \cite{1005.4840} or
the modulation mechanism \cite{1202.4580}, and the quintessence models for dark
energy \cite{1203.6655}. However, there is no concretely analysis of the
condition for generating the poly-instanton corrections in these papers.
Actually, within the most simple constructions, the poly-instanton effects in
these papers should be absent, as pointed out in \cite{1205.2485}.

On the other hand, most of particle physics models and multi-field inflationary
models in type IIB theory require at least four K\"ahler moduli. For
particle physics models, the moduli stabilisation seems to be more complicated,
since
one needs to consider the tension between non-perturbative effects 
and chirality\cite{0711.3389} also together with the $D$-term problem
\cite{1110.3333}. Thus the moduli stabilisation in this case must be done for an
explicit Calabi-Yau threefold, for example \cite{0811.4599,1110.3333}. For the
double-field inflation model in \cite{1005.4840,1202.4580}, we do not need to
worry about the tension between non-perturbative effects and chirality and
either the $D$-term problem, but one needs a K3-fibred Calabi-Yau threefold with
two del Pezzo surfaces, which is not explicitly presented in these papers
either. 

In the present paper, we will address the issue of moduli stabilisation for a
K3-fibred
Calabi-Yau threefold with {\sl four K$\ddot{\textrm{a}}$hler moduli}, which is a
crucial ingredient for realistic particle physics models or for the multi-field
inflation model construction in the type IIB orientifold framework. Within the
stabilisation mechanism presented here, all of the K$\ddot{\textrm{a}}$hler
moduli will be fixed by the non-perturbative effects, including poly-instanton
corrections, and $\alpha^\prime$-corrections. Moreover this procedure will not
be spoiled by the string-loop corrections to K\"ahler potentential. We will 
concretely check all of the conditions for generating non-perturbative
corrections, especially poly-instanton corrections, to the superpotential for an
explicit toric K3-fibred Calabi-Yau threefold, making sure that this kind of
superpotential can indeed be generated in the type IIB orientifolds, at least in
the simple cases where the background fluxes are ignored.

The paper is organized as follows. In section 2, we briefly review some
definitions relevant to $N=1$ type IIB orientifold compactifications with fluxes
as well as the general structure of LVS. In section 3, we will analyse the
conditions for the presence of non-perturbative corrections, including
poly-instanton corrections, to the superpotential in detail. First, we will
reiterate the neutral zero-mode and charged zero-mode issues of instantons,
then we will briefly mention the tools for calculating the cohomology group as
well as its splitting properties. Then we will present an explicit
Calabi-Yau manifold with an appropriate choice of orientifold action, which can
satisfy all of the conditions for generating the expected corrections to the
superpotential. In section 4, we will systematically present the procedure of
moduli stabilisation in type IIB orientifolds, in which the general compact
Calabi-Yau manifolds have the same structure as the one presented in section
3.1. Here we will find that we can not get a consistent result by using the
minimal superpotential at the end of section 3.1. But using instead the
racetrack superpotential, everything can be made consistent. We will also
discuss that the string-loop effects will not spoil this procedure. Finally in
section 5 we present our conclusions followed by an appendix providing the list
of K3 fibrations with del Pezzo and W-surface.

\section{Brief review of LARGE volume scenario}
In the framework of type IIB string theory compactified on a Calabi-Yau
threefold
with RR and NS-NS 3-form fluxes(see \cite{0509003} for reviews), we need
orientifold planes to reduce the supersymmetry, such that the low-energy
effective theory is a $N=1$ supergravity theory. The presence of orientifold
planes is also crucial to cancel the RR tadpoles \cite{0105097}. Depending on
the
transformation properties of the holomorphic three form $\Omega$ on Calabi-Yau
threefold, there are two different symmetry operations $\mathcal{O}$ to generate
the orientifold planes. Here we take the choice as follows, which can generate
O3/O7-planes
\begin{equation}
 \mathcal{O}=(-1)^{F_L}\Omega_p \sigma^\ast, \quad \sigma^\ast 
\Omega=-\Omega, \quad \sigma^\ast J=J
\label{eq:op}
\end{equation}
where $F_L$ is the spacetime fermion  number in the left-moving sector,
$\Omega_p$ denotes the world-sheet parity, $\sigma^\ast$ is the pull-back of
involution $\sigma: x_i \mapsto -x_i$, and the fixed point loci are defined as
O3/O7-planes. Note that since $\sigma$ is a holomorphic involution, the
cohomology groups $H^{(p,q)}$ split into two eigenspaces under the action
$\sigma^\ast$, namely $H^{(p,q)}=H_+^{(p,q)}\oplus H_-^{(p,q)}$. The
transformation properties for all of fields in type IIB
supergravity can be found in \cite{0403067}. 

The presence of O3/O7-planes wrapping a divisor gives rise to tadpoles for the
RR
form, which can be canceled by introducing suitable D3/D7-branes. Because we
only discuss moduli stabilisation, we assume for simplicity that there
are no gauge fluxes on the D7-brane. In this case we can avoid the Freed-Witten
anomaly \cite{9907189} by choosing suitable background B-field, and the
cancellation conditions read as
\begin{eqnarray}
 \sum_i N_i([D7_i]+[D7_i]^\prime)&=&8[O7], \nonumber \\ 
 N_{D3}+\frac{1}{2}N_{\textrm{flux}}-\frac{1}{4}N_{O3}&=&\frac{\chi(X)}{24}.
 \label{eq:tad}
\end{eqnarray}
Here $[D7]$ and $[O7]$ denote the divisors wrapped by D7-branes and O7-planes
respectively, $[D7]^\prime$ denotes the orientifold image of $[D7]$, $N_{D3}$ is
the
net number of D3-brane, namely the difference between the number of D3-branes
and the number of $\overline{D3}$-branes and
$N_{\textrm{flux}}=\frac{1}{(2\pi)^4
\alpha^{\prime2}}\int H_3 \wedge F_3$ and $\chi(X)$ is the Euler characteristic
of the elliptically fibered Calabi-Yau fourfold X. This framework can be viewed
as a limit of F-theory compactified on X, whose Euler characteristic is related
to the D7-branes and the O7-planes in type IIB theory as follows
\cite{0805.1573}
\begin{equation}
 2\chi(X)=\chi_o([D7])+4\chi([O7]).
\end{equation}
The modified Euler characteristic $\chi_o([D7])$ is defined as follows
\begin{equation}
 \chi_o([D7]) \equiv 24 \int \Gamma_{\textrm{pure} D7},
\end{equation}
where $\Gamma_{\textrm{pure} D7}$ is the charge of a pure D7-brane wrapping
$[D7]$, and it is shown that $\chi_o([D7]+[D7]^\prime)=2\chi([D7])$, so the
tadpole cancellation condition for $C_4$-form in Eq.(\ref{eq:tad}) reduces to
\begin{equation}
 N_{D3}+\frac{1}{2}N_{\textrm{flux}}=\frac{1}{4}N_{O3}+\frac{1}{4}\chi([O7]).
 \label{eq:tad3}
\end{equation}
Therefore, we can set eight D7-branes right on top of the O7-plane to cancel the
tadpole for $C_8$ form and the condition for $C_4$ form can serve as a
consistency check that the number is indeed a integer.

The 4-dimensional effective action in type IIB theory, which is
compactified on a Calabi-Yau orientifold, can be expressed into the standard
$\textrm{N}=1$ supergravity form, namely the action can be completely determined
by a K$\ddot{\textrm{a}}$hler potential $K$, a holomorphic superpotential $W$
and a holomorphic gauge-kinetic coupling functions $f$. Here we only talk
about the terms which are concerned with moduli and we must stress that all of
the variables involved in the following are in Einstein frame\footnote{The
relation between string frame and Einstein frame can see the appendix of
\cite{1005.4840}}.  To leading order in $g_s$ and $\alpha^\prime$, the
K$\ddot{\textrm{a}}$hler potential in Einstein frame is given as
\cite{0204254,0403067}
\begin{equation}
 K=-2 \log\left[\mathcal{V}+\frac{\hat{\xi}}{2}\right]-\log\left[-i \int
\Omega\wedge\bar{\Omega}\right]
 -\log\left[-i\left(\tau-\bar{\tau}\right)\right],
 \label{eq:KP}
\end{equation}
where $\mathcal{V}$ is the volume of the Calabi-Yau manifold and the
$\hat{\xi}$-term, which comes from $\alpha^\prime$-corrections, is expressed as
\begin{equation}
 \mathcal{V}=\frac{1}{6}\kappa_{\alpha\beta\gamma}t^\alpha t^\beta t^\gamma,
\qquad \hat{\xi}=-\frac{\zeta(3)\chi(\textrm{M})}{2(2\pi)^3 g_s^{3/2}}
 \label{eq:xi}
\end{equation}
respectively. In the previous equations, we have used the fact that the string
coupling $g_s=e^\phi$, and $\zeta(3)\approx 1.202$ is the approximate value of
Riemann $\zeta$-function, $\chi(\textrm{M})$ is the Euler characteristic of the
Calabi-Yau manifold. In order to perform the LARGE volume scenario, we must
require that $\hat{\xi}>0$ \cite{0502058}, namely $h^{2,1}>h^{1,1}$. 
$\kappa_{\alpha\beta\gamma}$ denotes the intersection numbers of the Calabi-Yau
manifold, and $t^\alpha$ denotes the coefficients of K$\ddot{\textrm{a}}$hler
form
on the basis of $H^{(1,1)}_{+}$. Furthermore, the holomorphic three-form
$\Omega$ in Eq.(\ref{eq:KP}) only depends on the complex structure moduli, and
the dilaton $\tau$ and K$\ddot{\textrm{a}}$hler moduli $T_\alpha$ take the
following definitions:
\begin{eqnarray}
 \tau&=&C_0+i e^{-\phi}, \nonumber \\
 G^{a}&=&c^a-\tau b^a,   \nonumber \\
 \zeta_\alpha &=&-\frac{i}{\tau-\bar{\tau}}\kappa_{\alpha bc}G^b (G-\bar{G})^c,
\quad a,b,c=1,\dots,h^{1,1}_{-} \, , \nonumber \\
 T_\alpha &=&\frac{1}{2}\kappa_{\alpha\beta\gamma}t^\beta
t^\gamma+i\rho_\alpha-\zeta_\alpha, \quad \alpha,\beta,\gamma=1,\dots,
h^{1,1}_{+}\,.
 \label{eq:KM}
\end{eqnarray}
 $c^a$ and $b^a$ are the coefficients of $C_2$ and $B_2$ on the basis of
$H^{(1,1)}_{-}$ respectively, $\rho_\alpha$ is the coefficient of $C_4$ on the
basis of $H^{(2,2)}_{+}$. Note that the basis of $H^{(1,1)}$ and $H^{(2,2)}$ are
dual to each other and $\pm$ denotes the two eigenspaces of splitting
cohomology groups $H^{(p,q)}$ under the involution $\sigma^\ast$. Under our
constructions, we can always set $h^{1,1}_{-}=0$, so that the
K$\ddot{\textrm{a}}$hler moduli can be simplified as 
\begin{equation}
 T_\alpha=\frac{1}{2}\kappa_{\alpha\beta\gamma}t^\beta t^\gamma+i \rho_\alpha.
\end{equation}
Note that $\tau_\alpha\equiv\frac{1}{2}\kappa_{\alpha\beta\gamma}t^\beta
t^\gamma$ can be viewed as the volume of divisor $D_\alpha \in
H_4(M,\mathbb{Z})$.

Ignoring gauge sectors, for orientifolds with $h^{1,1}_{-}=0$, the
superpotential $W$ in the perturbative theory was shown to be the
Gukov-Vafa-Witten
superpotential: \cite{9906070}
\begin{equation}
 W=\int \Omega \wedge G_3, \quad G_3=F_3-\tau H_3\, .
 \label{eq:GVW}
\end{equation}
Note that the superpotential is independent of the K$\ddot{\textrm{a}}$hler
moduli, and the K$\ddot{\textrm{a}}$hler potential possesses the well-known
no-scale structure. Thus the scalar potential of $N=1$ supergravity
\begin{equation}
 V=e^K\left[K^{I\bar{J}}D_I W \bar{D}_{\bar{J}} \bar{W}-3|W|^2\right]
 \label{eq:Pot}
\end{equation}
is positive definite and only depends on the dilaton and the complex structure
moduli.
We can fix both of them by solving
\begin{equation}
 D_a W\equiv \partial_a W + K_a W =0.
\end{equation}
Here $a$ runs over the  dilaton and complex structure moduli. Actually this
moduli fixing can be
done for appropriate choice of the fluxes \cite{0509003}, and from now on we
denote the value of $W$ following this step as $W_0$.

Therefore, in order to fix K$\ddot{\textrm{a}}$hler moduli, we must introduce
some non-perturbative effect, such as instanton corrections or gaugino
condensation effects to the superpotential as suggested by \cite{0502058}. In
the next section we will discuss the condition for generating such effects in
details.

\section{Superpotential with non-perturbative effects}
The superpotential with non-perturbative corrections, including instanton
effects,
poly-instanton effects or gaugino condensation takes the following form
\begin{eqnarray}
 W&=&W_0+A_i \exp(-a_i T_i+A_j e^{-2\pi T_j}) \nonumber \\
 &=&W_0+A_i e^{-a_i T_i}+A_i A_j e^{-a_i T_i-2\pi T_j}+\cdots,
\end{eqnarray}
where the instantons or the gaugino condensation are supported by the divisors
$D_i$, the poly-instantons are supported by the divisors $D_j$, the
corresponding K\"ahler moduli are $T_i, T_j$ respectively, and $A_{i},A_{j}$ are
one-loop determinants, which depend
 on complex structure moduli. 
$a_i=\frac{2\pi}{N}, N \in \mathbb{Z}_{+}$, where for D-brane instantons $N=1$,
while for gaugino condensation the value of $N$ depends on the rank of the gauge
group. In this section we will systematically analyse the condition for
generating such kind of
superpotential, using the methods in
\cite{0404257,1009.5386,1205.2485}.

Each BPS D-brane instanton is 1/2 BPS, and thus locally breaks 4 out of 8 the
supersymmetries. These broken supersymmetries manifest themselves in the volume
of the instanton as Goldstinos, namely as fermionic zero modes. They are
conventionally denoted by $\theta^\alpha$ and $\bar{\tau}_{\dot{\alpha}}$.
Depending on the divisor wrapped by the instanton, some other neutral zero modes
may also be present. In addition to the geometric Calabi-Yau background,
various other ingredients, such as branes, orientifolds and fluxes, maybe change
the spectrum of zero modes(See \cite{0902.3251} for a brief review on D-brane
instanton). In this paper we only consider the geometric Calabi-Yau background
as in \cite{0404257,1009.5386,1205.2485}. The general structure for the neutral
zero modes of an $O(1)$ instanton is showed in the table \ref{tab:ins}, in
which $\gamma_\alpha$ and $\bar{\gamma}_{\dot{\alpha}}$ denote the Wilson line
Goldstinos; $\chi_\alpha$ and $\bar{\chi}_{\dot{\alpha}}$ denote the deformation
Goldstinos.
\begin{table}[h]
\begin{center}
\begin{tabular}{c||c|c|c|c|c|c}
 \textrm{Zero Modes} &
$(X_\mu,\theta^\alpha)$&$\bar{\tau}_{\dot{\alpha}}$&$\gamma_\alpha$&$(\omega,
\bar{\gamma}_{\dot{\alpha}})$&$\chi_\alpha$&$(c,\bar{\chi}_{\dot{\alpha}})$\\
 \hline
 \textrm{Number} &$h^{0,0}_{+}(D)$&$h^{0,0}_{-}(D)$& $h^{1,0}_{+}(D)$ &
$h^{1,0}_{-}$(D) &
$h^{2,0}_{+}(D)$ & $h^{2,0}_{-}(D)$ 
\end{tabular}
\caption{\label{tab:ins}Neutral zero mode structure for an $O(1)$-instanton
wrapping a
divisor $D$ \cite{1009.5386}}
 \end{center}
\end{table}
If the instanton contributes to the holomorphic superpotential $W$, the
anti-holomorphic zero modes have to be removed, namely $h^{n,0}_{-}=0,
n=0,1,2,\cdots$, and they should be no more other zero modes, i.e.
$h^{1,0}(D)=h^{2,0}(D)=0$. For the contribution from gaugino condensation, the
condition of the divisor is the same as before, actually in this case we have an
ordinary gauge instanton for a Sp(2N) or SO(N) gauge group. Considering these
constraints and the realization of the LARGE volume scenario, the divisor which
supports an instanton or gaugino condensation has to be a del-Pezzo surface
$dP_n$,
since they are arbitrarily contractible to a point without affecting the rest of
geometry on a Calabi-Yau threefold \cite{0111068}, since a brane wrapping such
a surface has no adjoint  matter and no extra fermionic modes. For the
contribution from poly-instanton, the former Wilson line Goldstinos of the E1
instanton, for an E3 instanton, can arise from either Wilson line or deformation
Goldstinos, which are counted by $h^{1,0}_{+}(D)+h^{2,0}_{+}(D)$. So we can
summarize the conditions for the zero mode structure of an instanton and a
poly-instanton contribution for the superpotential in the following table
\ref{tab:con}.
\begin{table}[h]
\begin{center}
 \begin{tabular}{c||c|c|c|c}
  &$h^{0,0}_{+}(D)$& $h^{1,0}_{+}(D)$ & $h^{2,0}_{+}(D)$ & $h^{n,0}_{+}(D)$\\
  \hline
  Instanton & 1 & 0 &0 & 0 \\
  Poly-Instanton & 1 & 1 or 0 & 0 or 1 & 0
 \end{tabular}
 \caption{\label{tab:con}Neutral zero mode structure of an instanton and a
poly-instanton wrapping on a divisor $D$ in order to contribute to the
superpotential
\cite{1205.2485}.}
 \end{center}
 \end{table}
Actually as argued in \cite{1205.2485}, only the divisor that admits a single
complex
Wilson line Goldstino can really support a poly-instanton correction. They call
it a W-surface, which is charactered by $(h^{0,0},h^{1,0},h^{2,0})=(1,1,0)$.

Because of the presence of D7-branes, so in addition of open strings going from
the instanton to itself, which gives rise to neutral zero modes, there are also
open strings going from instanton to D7 branes, which generates charged zero
modes transforming in the fundamental or anti-fundamental representation of
D7-brane gauge group. These charged zero modes can couple to the matter fields
on D7-brane, after integrating over these zero modes can induce some effective
operators involving matter fields in the low energy effective theory. It implies
that in order to know exactly the physical charged zero modes, we need to
understand the structure of Yukawa couplings in our compactification. And that
is beyond the scope of this paper, so we will simply require the absence of
charged zero modes on the instanton.

Consider an instanton A and a background D7-brane wrapping different divisors,
that intersect over the curve $\mathcal{C}=[A]\cdot [D7]$. The spectrum of
charged zero modes from the open string going from the instanton A and the 
D7-brane originates from the cohomology group \cite{0307245}
\begin{equation}
 (\alpha,\bar{\beta})\in(H^0(\mathcal{C},K^{1/2}_{\mathcal{C}}),H^1(\mathcal{C},
K^{1/2}_{\mathcal{C}}))\, ,
\end{equation}
where $\alpha$ and $\bar{\beta}$ denote the modes in the fundamental and
anti-fundamental representation of the D7-brane gauge group, and
$K_{\mathcal{C}}$ stands for the anticanonical bundle of $\mathcal{C}$. We can
ensure there are no charged zero modes at least in the following two cases
\cite{1009.5386}:
\begin{itemize}
 \item $\mathcal{C}=0$, namely there is no intersection between the instanton
and  the D7-brane,
 \item $\mathcal{C}=\mathbb{P}^1$.
\end{itemize}
From the above analysis, almost all of the conditions have been translated into
the
language of cohomology. Hence in the following we will briefly discuss the tools
to
calculate the cohomology group and its splitting under the orientifold
involution. First we have the usual holomorphic Euler characteristic of
the divisor $D$
\begin{equation}
 \chi(D,\mathcal{O}_D)=\sum_{i=0}^{2}(-1)^i h^{i,0}(D),
 \label{eq:ec}
\end{equation}
and it is easy to compute using the Riemann-Roch formula:
\begin{equation}
 \chi(D,\mathcal{O}_D)=\int_D Td(TD)\, .
\end{equation}
$Td(TD)$ is the Todd classes of tangent bundle to $D$. On the other hand,
recalling the splitting
$H^i(D,\mathcal{O}_D)=H^i_{+}(D,\mathcal{O}_D)\oplus H^i_{-}(D,\mathcal{O}_D)$,
we immediately have $h^{i,0}(D)=h^{i,0}_{+}(D)+h^{i,0}_{-}(D)$, and we also have
the Lefschetz's equivariant genus for the orientifold involution $\sigma$ as
\begin{equation}
 \chi^\sigma(D,\mathcal{O}_D)=\sum_{i=0}^{2}(-1)^i
(h^{i,0}_{+}(D)-h^{i,0}_{-}(D))\, .
 \label{eq:leg}
\end{equation}
On the other hand, we can easily compute the Lefschetz's equivariant genus from
the
Lefschetz fixed point theorem, where one can see some details of the theorem in
the appendix of \cite{1009.5386},
\begin{equation}
 \chi^\sigma(D,\mathcal{O}_D)=\frac{1}{4} N_{O3}-\frac{1}{4}
\int_{\mathcal{C}^\sigma} [D]\, ,
\end{equation}
where $N_{O3}$ is the number of isolated fixed points on $D$,
$\mathcal{C}^\sigma=[O7]\cap D$ are the fixed curves on $D$, and $[D]\in
H^2(\textrm{M})$ denotes the Poincare dual to the divisor $D$. For the
equivariant
Betti number,  a similar theorem applies, which leads to
\begin{equation}
 L^\sigma(\textrm{M})=\sum_{i=0}^4
(-1)^i(b^i_{+}-b^i_{-})=N_{O3}+\chi(\mathcal{C}^\sigma)\, .
\end{equation}
In most of the cases, using the above equations, we can
determine all equivariant cohomology classes. For other cases, we can employ the
tools presented in \cite{1010.3717} for the computation of line bundle
cohomology over toric varieties, where we obtain the polytope information, which
is crucial to construct the Calabi-Yau threefold, by the help of PALP package
\cite{0204356}.

Using the tools mentioned above, searching through the 158 examples of four
dimensional reflexive lattice polytopes presented in the \cite{1107.0383}, which
admit a K3-fibred Calabi-Yau hypersurface with four K$\ddot{\textrm{a}}$hler
moduli where at least one of them is a del Pezzo surface, we find that 23 of
them can also admit one W-surface. We will present all of them in the
appendix.

Next we will pick one of the reflexive lattice polytopes, namely No.3 in the
appendix, to show explicitly that the del Pezzo surfaces and W-surfaces can
indeed support an instanton and a poly-instanton, which contribute to
superpotential.
\subsection{An explicit example}
The toric ambient space can be defined by homogeneous coordinates and their
equivalence relations, which are all encoded in the following weight matrix:

 \begin{center}
  \begin{tabular}{c c c c c c c c |c}
   $x_1$&$x_2$&$x_3$&$x_4$&$x_5$&$x_6$&$x_7$&$x_8$&$D_H$ \\
   \hline
   2 & 1& 6& 1& 2& 0& 0& 0& 12\\
   2 & 1& 6& 0& 1& 0& 2& 0& 12\\
   2 & 0& 6& 1& 1& 0& 0& 2& 12\\
   1 & 0& 3& 0& 1& 1& 0& 0& 6
  \end{tabular}
 \end{center}

We can show that the surface of this reflexive lattice polytope admit 4 maximal
triangulations, considering that we may construct a fan from each triangulation,
so we can obtain several toric varieties for this weight matrix. Actually the
different triangulations are not isolated from each other, they maybe related to
each other via flop transitions. Here we stick to one of the triangulations,
which
is encoded in the following Stanley-Reisner(SR) ideal
\begin{equation}
 \textrm{SR}=\{x_2 x_4, x_2 x_5, x_4 x_5, x_5 x_6, x_2 x_7, x_4 x_8, x_1 x_3 x_6
x_7, x_1 x_3 x_6 x_8, x_1 x_3 x_7 x_8\}\, .
\end{equation}
The Calabi-Yau hypersurface in this toric ambient space has Hodge number
$(h^{1,1},h^{1,2})=(4,70)$ and Euler characteristic $\chi=-132$. From the
SR-ideal and the weight matrix, we can calculate the triple intersection numbers
for a basis of divisor classes of the toric variety on Calabi-Yau. For
simplifying the expression of the volume, we choose the following basis
$(\eta_1,\eta_2,\eta_3,\eta_4)=(D_2,D_4,D_5,D_5+6D_4+3D_7)$, where divisors
$D_i\equiv\{x_i=0\}$. The triple intersection numbers can be expressed in the
following polynomial:
\begin{equation}
 I_3=\eta_1^3+\eta_2^3+18\eta_3 \eta_4^2\, .
\end{equation}
The generators $\mathcal{C}_i$ of the Mori cone of the toric variety are
\begin{equation}
 \int_{C_i} D_j = \left(\begin{array}{c c c c}
                         -1 & 0 & 0 & 0 \\
                          0 & 0 & 1 & -2 \\
                          1 & 0 & 0 & 3 \\
                          0 &-1 & 0 & 0 \\
                          0 & 1 & 0 & 3
                        \end{array}
\right).
\end{equation}
We know that the K$\ddot{\textrm{a}}$hler cone and Mori cone are dual to each
other, and from this we can get the generators of the K$\ddot{\textrm{a}}$hler
cone as follows:
\begin{eqnarray}
 \Gamma_1&=&\eta_3 \, ,\nonumber \\
 \Gamma_2&=&2\eta_3+\eta_4\,, \nonumber \\
 \Gamma_3&=&-3\eta_2+2\eta_3+\eta_4\, , \nonumber \\
 \Gamma_4&=&-3\eta_1-3\eta_2+2\eta_3+\eta_4 \, ,\nonumber \\
 \Gamma_5&=&-3\eta_1+2\eta_3+\eta_4\, .
\end{eqnarray}
This  explicitly shows  that the polytope is non-simplicial since $i$ runs from
one
to five instead of four. Next we write the K$\ddot{\textrm{a}}$hler form in the
basis
of $\{\eta_i\}$ and $\{\Gamma_i\}$ respectively
\begin{equation}
 J=\sum_{i=1}^4 t_i \eta_i=\sum_{i=1}^5 r_i \Gamma_i \quad \textrm{with} \quad
r_i>0\, ,
\end{equation}
where $r_i>0$ ensures that the stabilisation is within the
K$\ddot{\textrm{a}}$hler cone. Now we can obtain the volume form in terms of
two-cycle volume $t_i$,
\begin{equation}
 \mathcal{V}=\frac{1}{3!} \int J\wedge J\wedge J
=\frac{1}{6}\kappa_{ijk}t^i t^j t^k=\frac{1}{6}t_1^3+\frac{1}{6}t_2^3+9t_3 t_4^2
\end{equation}
and we can also express $t_i$ in terms of $r_i$, from which we can determine the
sign of $t_i$ and the linear combination of them.
\begin{eqnarray}
 t_1&=&-3r_4-3r_5 <0, \nonumber \\
 t_2&=&-3r_3-3r_4 <0,   \nonumber \\
 t_3&=&r_1+2r_2+2r_3+2r_4+2r_5 >0 \, ,\nonumber \\
 t_4&=&r_2+r_3+r_4+r_5 >0 \, .
 \label{eq:Kc}
\end{eqnarray}
Defining the volumes $\tau_i$ of the four-cycle $D_i$,
\begin{equation}
 \tau_i=\frac{1}{2}\int_{D_i} J\wedge J=\frac{1}{2}\kappa_{ijk}t^j t^k\, ,
\end{equation}
we find that
\begin{eqnarray}
 \tau_1&=&\frac{1}{2}t_1^2\, , \nonumber \\
 \tau_2&=&\frac{1}{2}t_2^2 \, ,\nonumber \\
 \tau_3&=&9t_4^2\, , \nonumber \\
 \tau_4&=&18t_3 t_4\, .
\end{eqnarray}
Taking into account the K$\ddot{\textrm{a}}$hler cone condition (\ref{eq:Kc}),
we can
rewrite the volume form in terms of four-cycle's volumes\footnote{Note that if
we choose a basis including the W-surface divisor, the `minus'-part in the
volume form will be similar to the examples in \cite{1205.2485}}
\begin{equation}
 \mathcal{V}=\frac{1}{6}\sqrt{\tau_3}\tau_4-\frac{\sqrt{2}}{3}\tau_1^{3/2}-\frac
{\sqrt{2}}{3}\tau_2^{3/2}\, .
 \label{eq:Vol}
\end{equation}

Next we will analyze the properties of the divisors $D_i,i=1,\cdots,8$, and show
that there are two del Pezzo surfaces, one W-surface, and also the Calabi-Yau
hypersurface is indeed K3-fibred. First of all, after computing the Hodge
diamonds, we find that both $D_2$ and $D_4$ have the topological data of $dP_8$,
namely $(h^{0,0},h^{1,0},h^{2,0},h^{1,1})=(1,0,0,9)$ and $\chi=11$, and also the
following triple intersection structures with the other divisors:
\begin{center}
\begin{tabular}{c|c c c c c c c c}
      &$D_1$&$D_2$&$D_3$&$D_4$&$D_5$&$D_6$&$D_7$&$D_8$ \\
 \hline
 $D_2^2$& -1  & 1 & -3 & 0 & 0 & -1 & 0 & -2 \\
 $D_4^2$& -1  & 0 & -3 & 1 & 0 & -1 & -2 & 0
\end{tabular}
\end{center}
The divisors $D_2$ and $D_4$ have triple self-intersections $D_2^3=1,D_4^3=1$,
any other intersection numbers are either vanishing or negative. This reads
\begin{equation}
 \int_{\mathcal{C}=D_i\cap S} c_1(S)=\int_M D_i\wedge S
\wedge(-c_1(\mathcal{N}_{S|M}))=-S^2\cdot D_i>0 \quad \forall
\mathcal{C}\neq\emptyset,
\end{equation}
where $D_i \neq S$. It is a necessary condition for the divisor $S$ to be a
rigid
and shrinkable divisor, i.e. del Pezzo surface. This confirms that these two
divisors should be $dP_8$. Actually we can also algebraically show that these
two
divisors are indeed $dP_8$. Following the procedure showed in \cite{Mayrhofer},
we can get the representation of these two divisors
\begin{eqnarray}
 D_2: \quad \begin{tabular}{c c c c c| c}
      $x_1$ & $x_3$ & $x_6$ & $x_7$ & $x_8$&$D_H|_{D_2}$ \\
      \hline
      2 & 6 & 0 & 2 & 2 & 12 \\
      1 & 3 & 1 & 0 & 2 &6
      \end{tabular}
\quad &\textrm{with}& \quad \textrm{SR}|_{D_2}=\{x_6 x_8, x_1 x_3 x_6 x_7, x_1
x_3 x_7 x_8\}, \quad \textrm{and} \nonumber \\
D_4: \quad \begin{tabular}{c c c c c| c}
      $x_1$ & $x_3$ & $x_6$ & $x_7$ & $x_8$&$D_H|_{D_4}$ \\
      \hline
      2 & 6 & 0 & 2 & 2 & 12 \\
      1 & 3 & 1 & 2 & 0 &6
      \end{tabular}
\quad &\textrm{with}& \quad \textrm{SR}|_{D_4}=\{x_6 x_8, x_1 x_3 x_6 x_7, x_1
x_3 x_7 x_8\}.
\end{eqnarray}
Hence $D_2$ and $D_4$ are both $dP_8$-surface.

As a next step, we find out the K3 divisor. We can easily compute that
$\int_{D_5} c_1(D_5)\wedge i^\ast D_i=-D_5^2 D_i=0$ and $\int_{D_5} i^\ast
c_2(\textrm{M})=\int_{D_5}
(10\eta_1^2-28\eta_1\eta_2+18\eta_2^2-64\eta_1\eta_3-28\eta_2\eta_3+\frac{8}{3}
\eta_3^2-8\eta_1\eta_4+8\eta_2\eta_4+6\eta_3\eta_4+\frac{4}{3}\eta_4^2)=24>0$,
the main theorem of \cite{Oguiso} implies that this Calabi-Yau threefold is a K3
fibration over $\mathbb{P}^1$ with typical fibre $D_5$. The explicit
computation of the Hodge diamond of $D_5$ leads to
$(h^{0,0},h^{1,0},h^{2,0},h^{1,1})=(1,0,1,20)$ and $\chi=24$, which is exactly
the topological data of K3 surface. And we can also compute the Hodge
diamond of $D_6$, $(h^{0,0},h^{1,0},h^{2,0},h^{1,1})=(1,1,0,4)$ and $\chi=2$,
which is exactly the topological data of W-surface.

Finally let us analyse the splitting properties of the cohomology group under
the
orientifold involution $\sigma$. We restrict that the orientifold involution
$\sigma$ just flips the sign of one homogeneous coordinates, i.e. $\sigma:
x_i\mapsto -x_i$, and we find that there are three inequivalent involutions
$\sigma:\{x_2\mapsto -x_2,x_4\mapsto -x_4, x_5\mapsto -x_5\}$, such that the
W-surface has the appropriate splitting properties and
$h^{1,1}_{-}(\textrm{M})=0$. In the following we take the involution $\sigma:
x_5\mapsto -x_5$ as an example. Following the algorithm presented in
\cite{Mayrhofer}, we obtain the following fixed point set of the ambient space
\begin{equation}
 \{\textrm{Fixed}\}|_{x_5\leftrightarrow
-x_5}^{\textrm{Ambient}}=\{D_5,D_2,D_4,D_1\cdot D_3\cdot D_6, D_6\cdot D_7\cdot
D_8\}\, .
\end{equation}
From the generic hypersurface equation, only the following subset of the fixed
point intersects the invariant hypersurface
\begin{equation}
 \{\textrm{Fixed}\}|_{x_5\leftrightarrow
-x_5}^{\textrm{CY}}=\{D_5,D_2,D_4,D_6\cdot D_7 \cdot D_8\}\, .
\end{equation}
Hence we have three O7-planes wrapping the divisors $D_2, D_4, D_5$ respectively
and two O3-planes, since the intersection number $D_6\cdot D_7\cdot D_8=2$. And
we need to put eight D7-branes right on top of each O7-plane to cancel the
D7-brane tadpole. Since the D7-branes are wrapping on the del Pezzo surfaces
$D_2$ and $D_4$, which are pointwise invariant under the involution $\sigma$, so
the only possible non-perturbative corrections to superpotential are from
gaugino condensations on the D7-branes instead of D-brane instantons. Meanwhile
applying Eq.(\ref{eq:tad3}), the contribution to D3-brane tadpole is
\begin{equation}
 N_{D3}+\frac{1}{2}N_{\textrm{flux}}=\frac{1}{4}(2+11+11+24)=12\, ,
\end{equation}
which is indeed integer as required. The splitting Hodge number of the
del-Pezzo surfaces and W-surface under the involution $\sigma$ is as follows:
\begin{eqnarray}
 D_2: (h^{0,0},h^{1,0},h^{2,0},h^{1,1})&=&(1_{+},0,0,9_{+})\, , \nonumber \\
 D_4: (h^{0,0},h^{1,0},h^{2,0},h^{1,1})&=&(1_{+},0,0,9_{+})\, , \nonumber \\
 D_6: (h^{0,0},h^{1,0},h^{2,0},h^{1,1})&=&(1_{+},1_{+},0,4_{+})\, .
\end{eqnarray}
So we have the correct topological data for the neutral zero modes. Furthermore,
we can read from the SR-ideal that the divisors supporting D7-branes, which lie
on top of O7-planes, do not intersect with each other, and also there is no
intersection between $D_5$ and $D_6$. Furthermore, the intersection between
$D_2$ and $D_6$ is a surface with genus 1, which can be determined to be $T^2$
according to the classification theorem of the closed surface. We can summarize
all this topological and geometrical information in the table \ref{tab:info}.
\begin{table}[h]
 \begin{center}
  \begin{tabular}{c|c|c}
   Divisor& $(h^{0,0},h^{1,0},h^{2,0},h^{1,1})$& Intersection Curves\\
   \hline \hline
   $D_2=dP_8$ &  $(1_{+},0,0,9_{+})$ & $D_6:C_{g=1}$ \\
   $D_4=dP_8$ &  $(1_{+},0,0,9_{+})$ & $D_6:C_{g=1}$ \\
   $D_5=K3$ &  $(1_{+},0,1_{+},20_{+})$ & Null\\
   $D_6=W$ &  $(1_{+},1_{+},0,4_{+})$  & $D_2: C_{g=1},\quad D_4: C_{g=1}$
  \end{tabular}
 \end{center}
 \caption{\label{tab:info}Divisors with topological and geometrical
information.}
\end{table}

Since W intersects the D7-branes over a $T^2$ and
$h^{\ast}(T^2,\mathcal{O})=(1,1)$, there will be extra vector-like zero modes.
If there is a non-trivial Wilson line on $T^2$, these zero modes can pair up and
become massive \cite{0811.2936}. For this purpose, one must have the freedom to
turn on an additional gauge bundle on the divisor $[D7]$, whose restriction on
the intersection curve $T^2$ is a non-trivial Wilson line. As argued in
\cite{1205.2485}, an additional gauge bundle which  is supported only on
2-cycles $C_i \subset [D7]$, which are topological trivial in M but do intersect
with the curve $T^2$, allows one to avoid these extra zero modes. Considering
that both $D_2$ and $D_4$ have more 2-cycles than the Calabi-Yau M, since
$h^{1,1}(D_2)=h^{1,1}(D_4)>h^{1,1}(M)$,  they must therefore exist such trivial
2-cycles.

Finally after checking all of constraints, we can ensure that the following
superpotential can indeed be generated by the gaugino condensations on D7-branes
and the poly-instanton effects
\begin{equation}
 W=W_0+A_1 e^{-a_1 T_1}+A_1 A_6e^{-a_1 T_1}e^{-2\pi(T_3-T_1-T_2)}+A_2 e^{-a_2
T_2}+A_2 A_6e^{-a_2 T_2}e^{-2\pi(T_3-T_1-T_2)}\, .
 \label{eq:W}
\end{equation}
Note that $D_6=\eta_3-\eta_1-\eta_2$ and the K\"ahler moduli $T_i,i=1,2,3$ in
the previous equation is associate to the volume modulus of basis divisors
$\eta_i$ respectively.

\section{Moduli stabilisation}
In this section, we will discuss the K\"ahler moduli stabilisation using the
superpotential (\ref{eq:W}) and the K$\ddot{\textrm{a}}$hler potential
(\ref{eq:KP}) with the general volume form
\begin{equation}
 \mathcal{V}=\alpha\sqrt{\tau_3}\tau_4-\beta_1 \tau_1^{3/2}-\beta_2 \tau_2^{3/2}
 \label{eq:GK}
\end{equation}
which has the same volume structure as the explicit example (\ref{eq:Vol}) and
$\alpha, \beta_1, \beta_2$ are some real constants.

As suggested in \cite{0502058}, the scalar potential (\ref{eq:Pot}) in the
LARGE volume scenario can be divided into three parts $V_{np1}, V_{np2}$
and $V_{\alpha^\prime}$:
\begin{eqnarray}
 V&=&V_{np1}+V_{np2}+V_{\alpha^\prime} \, ,\nonumber \\
 V_{np1}&=&e^K K^{i\bar{j}}\partial_i W\partial_{\bar{j}}\bar{W}\, , \nonumber
\\
 V_{np2}&=&e^K K^{i\bar{j}}(\partial_i W
K_{\bar{j}}\bar{W}+\partial_{\bar{j}}\bar{W} K_i W)\, , \nonumber \\
 V_{\alpha^\prime}&=&e^K \left(K^{i\bar{j}}K_i
K_{\bar{j}}-3\right)|W|^2=e^K\left[3\hat{\xi}\frac{\hat{\xi}^2+7\hat{\xi}
\mathcal{V}+\mathcal{V}^2}{(\mathcal{V}-\hat{\xi})(2\mathcal{V}+\hat{\xi})^2}
|W|^2\right]\, .
\end{eqnarray}
Note that in the large volume limit we can ignore the
$\alpha^\prime$-corrections to the K$\ddot{\textrm{a}}$hler potential in the
expression of $V_{np1}$ and $V_{np2}$, and the $V_{\alpha^\prime}$ term reduces
to
\begin{equation}
 V_{\alpha^\prime}=e^K
\frac{3\hat{\xi}}{4\mathcal{V}}|W|^2=\frac{3\hat{\xi}}{4\mathcal{V}^3}|W|^2\, .
\end{equation}
We also expect the divisors $\eta_1$ and $\eta_2$ to be the small divisors
in the large volume limit.

In order to perform the calculation of the scalar potential, first of all we
need to get the K$\ddot{\textrm{a}}$hler metric and its inverse. The
K$\ddot{\textrm{a}}$hler metric is given by the following symmetric matrix
\begin{equation}
 K_{i\bar{j}}=\left(
\begin{array}{cccc}
 \frac{3\beta_1}{8 \mathcal{V} \sqrt{\tau _1}} & \frac{9\beta_1\beta_2\sqrt{\tau
_1}
   \sqrt{\tau _2}}{8\mathcal{V}^2} & -\frac{3\beta_1\sqrt{\tau _1}}{8\mathcal{V}
\tau _3} & -\frac{3\alpha\beta_1 \sqrt{\tau _1} \sqrt{\tau _3}}{4\mathcal{V}^2}
\\
 \frac{9\beta_1\beta_2\sqrt{\tau _1} \sqrt{\tau _2}}{8 \mathcal{V}^2} &
\frac{3\beta_2}{8\mathcal{V} \sqrt{\tau _2}} & -\frac{3\beta_2\sqrt{\tau
_2}}{8\mathcal{V}\tau _3} & -\frac{3\alpha\beta_2\sqrt{\tau _2} \sqrt{\tau
_3}}{4\mathcal{V}^2} \\
 -\frac{3\beta_1\sqrt{\tau _1}}{8\mathcal{V} \tau _3} &
-\frac{3\beta_2\sqrt{\tau
   _2}}{8\mathcal{V} \tau _3} & \frac{1}{4 \tau _3^2} & \frac{\alpha(\beta_1
\tau_1^{3/2}+\beta_2 \tau_2^{3/2})}{4\mathcal{V}^2 \sqrt{\tau _3}} \\
 -\frac{3\alpha\beta_1\sqrt{\tau _1} \sqrt{\tau _3}}{4\mathcal{V}^2} &
   -\frac{3\alpha\beta_2\sqrt{\tau _2} \sqrt{\tau _3}}{4\mathcal{V}^2} &
   \frac{\alpha(\beta_1
\tau_1^{3/2}+\beta_2 \tau_2^{3/2})}{4\mathcal{V}^2 \sqrt{\tau _3}} &
\frac{\alpha^2\tau _3}{2 \mathcal{V}^2} \\
\end{array}
\right),
\end{equation}
where we have used the expression of the volume in the large volume limit,
$\mathcal{V}=\alpha\sqrt{\tau_3}\tau_4$, and dropped the subleading terms
in orders of $\mathcal{V}$.

The inverse of K$\ddot{\textrm{a}}$hler metric in the large volume limits reads
as
\begin{equation}
 K^{i\bar{j}}=\left(
\begin{array}{cccc}
 \frac{8 \mathcal{V} \sqrt{\tau _1}}{3 \beta_1 } & 4 \tau _1 \tau _2 & 4 \tau _1
   \tau _3 & \frac{4 \mathcal{V} \tau _1}{\alpha  \sqrt{\tau _3}} \\
 4 \tau _1 \tau _2 & \frac{8 \mathcal{V} \sqrt{\tau _2}}{3 \beta_2 } & 4 \tau
   _2 \tau _3 & \frac{4 \mathcal{V} \tau _2}{\alpha  \sqrt{\tau _3}} \\
 4 \tau _1 \tau _3 & 4 \tau _2 \tau _3 & 4 \tau _3^2 & \frac{4 \beta_1 
   \sqrt{\tau _3} \tau _1^{3/2}}{\alpha }+\frac{4 \beta_2  \tau _2^{3/2}
   \sqrt{\tau _3}}{\alpha } \\
 \frac{4 \mathcal{V} \tau _1}{\alpha  \sqrt{\tau _3}} & \frac{4 \mathcal{V}
   \tau _2}{\alpha  \sqrt{\tau _3}} & \frac{4 \beta_1  \sqrt{\tau _3} \tau
   _1^{3/2}}{\alpha }+\frac{4 \beta_2  \tau _2^{3/2} \sqrt{\tau _3}}{\alpha }
   & \frac{2 \mathcal{V}^2}{\alpha ^2 \tau _3} \\
\end{array}
\right).
\end{equation}
So we can now calculate the scalar potential. In the LARGE volume scenario, we
can set $e^{-a_i \tau_i}\propto \mathcal{V}^{-1}, i=1,2$ and
$e^{-2\pi(\tau_3-\tau_1-\tau_2)}\propto \mathcal{V}^{-p}$, namely $e^{-2\pi
\tau_3}\propto \mathcal{V}^{-m_1-m_2-p}$ with $a_i=\frac{2\pi}{m_i}, m_i\in
\mathbb{Z}_{+}, i=1,2$. Then we stabilize the moduli order by order in
$1/\mathcal{V}$. All the possible orders in $\mathcal{V}$ in the expression of
scalar potential are
\begin{equation}
 \{\textrm{Possible Orders}\}=\{-3,-3-p,-3-2p,-4,-4-p,-4-2p,-5,-5-p,-5-2p\}\, .
 \label{eq:ord}
\end{equation}
Observing the structure of the inverse matrix of K$\ddot{\textrm{a}}$hler metric
and the structure of the superpotential, we can easily obtain that $\tau_3$ is
only involved in the term whose order includes $p$, and other terms are
independent of $\tau_3$. This observation is very important to determine the
value of $p$ in the end.

At the leading order, i.e. $\mathcal{O(V}^{-3})$, the scalar potential 
explicitly reads as
\begin{equation}
 V_{\mathcal{O(V}^{-3})}=\sum_{i=1}^2\left(\frac{8 a_i^2 A_i^2 \sqrt{\tau_i}
e^{-2 a_i\tau_i}}{3\beta_i \mathcal{V}}+\frac{4 a_i A_i W_0 \tau_i e^{-a_i
\tau_i} \cos\left(a_i\rho_i\right)}{\mathcal{V}^2}\right)+\frac{3 W_0^2
\hat{\xi}}{4 \mathcal{V}^3}\, .
\end{equation}
We start with minimising the scalar potential with respect to the axions
$\rho_1,\rho_2$, assuming that all involved parameters are real and positive. 
It is obvious that the minimal values lie at $a_i
\rho_i=(2k+1)\pi,k\in\mathbb{Z}$, namely
\begin{equation}
 \rho_i=\frac{1}{2}(2k+1)m_i, \quad k\in\mathbb{Z},i=1,2\, .
\end{equation}
Then we minimise scalar potential with respect to the K$\ddot{\textrm{a}}$hler
metric $\tau_1,\tau_2$, and the relevant derivatives equal to 0 can be reduced
as
\begin{eqnarray}
 \frac{\partial V}{\partial \tau_i}=0&:&\quad a_i A_i \mathcal{V}(1-4a_i
\tau_i)=3\beta_i W_0 e^{a_i\tau_i}(1-a_i\tau_i)\tau_i^{1/2}, \quad i=1,2
 \nonumber\\
 \frac{\partial V}{\partial \mathcal{V}}=0&:&\sum_{i=1}^2\left(-\frac{32a_i^2
A_i^2}{3\beta_i}\sqrt{\tau_i}e^{-2 a_i\tau_i}\mathcal{V}^2+32 a_i A_i W_0 \tau_i
e^{-a_i \tau_i} \mathcal{V}\right)=9W_0^2 \hat{\xi}.
\label{eq:der}
\end{eqnarray}
Considering that we require $a_i\tau_i\gg 1$ to reduce the higher instanton
corrections, the first equation in (\ref{eq:der}) reduces to
\begin{equation}
 4a_i A_i \mathcal{V}=3\beta_i W_0 e^{a_i\tau_i}\tau_i^{1/2}.
 \label{eq:id1}
\end{equation}
In order to obtain a consistent value of the volume, we can take a very natural
assumption that $a_1\tau_1=a_2\tau_2$. Taking this identity into the second
equation
of (\ref{eq:der}), we obtain
\begin{equation}
 2(\beta_1\tau_1^{3/2}+\beta_2\tau_2^{3/2})=\hat{\xi}\, .
 \label{eq:id2}
\end{equation}
Taking into account the assumption $a_1\tau_1=a_2\tau_2$, we get the
stabilisation
value of $\tau_1,\tau_2$ and $\mathcal{V}$:
\begin{eqnarray}
 a_i\langle\tau_i\rangle&=&\left(\frac{\hat{\xi}}{2J}\right)^{2/3},i=1,
2 \quad \textrm{with} \quad J=\sum_{i=1}^2\beta_i a_i^{-3/2}\, ,\nonumber\\
 \langle\mathcal{V}\rangle&=&\frac{3\beta_i W_0}{4a_i A_i}e^{a_i \langle\tau_i
\rangle}\langle\tau_i\rangle^{1/2} \quad \forall i=1,2 \, .
\end{eqnarray}
Next we consider the stabilisation of the fibration $\tau_3$. The first order
containing $T_3$ in the scalar potential is $\mathcal{O(V}^{-3-p})$, which reads
as
{\footnotesize
\begin{eqnarray}
 &&V_{\mathcal{O(V}^{-3-p})}\nonumber\\
 =&-&\frac{4 A_6
W_0e^{-(a_1-2\pi)\tau_1-(a_2-2\pi)\tau_2-2\pi\tau_3}}{\mathcal{V}^2}
\left[\tau_1\left((2\pi -a_1) A_1 e^{a_2 \tau_2}
+2\pi A_2e^{a_1\tau_1} \right) \right. \nonumber \\
&+&\tau_2 \left(2\pi A_1 e^{a_2\tau_2} +(2 \pi -a_2) A_2 e^{a_1\tau_1}\right)
 -\left. 2\pi \tau_3\left(A_1 e^{a_2 \tau _2} +A_2 e^{a_1 \tau_1}\right)\right]
 \cos\left((1-m_1-m_2)\pi+2\pi\rho_3\right)\nonumber\\
   &-&\frac{16 A_6
e^{-2\left((a_1-\pi)\tau_1+(a_2-\pi)\tau_2+\pi\tau_3\right)}}{
3\beta_1\beta_2\mathcal{V}}\left[a_1 A_1 \beta_2 \sqrt{\tau_1} e^{a_2 \tau _2}
\left((2 \pi -a_1)A_1 e^{a_2 \tau _2} +2\pi A_2 e^{a_1\tau_1}\right)\right.
 \nonumber \\
   &+&\left.a_2 A_2 \beta_1 \sqrt{\tau _2} e^{a_1 \tau_1} \left(2 \pi  A_1
e^{a_2 \tau_2} +(2 \pi -a_2) A_2 e^{a_1 \tau _1} \right)\right]\cos
\left((m_1+m_2)\pi-2\pi\rho_3\right)\, ,
\end{eqnarray}
}
where we have used the stabilised value of $\rho_i,i=1,2$ in this equation. Now
the scalar potential in the order of $\mathcal{V}^{-3-p}$ can be rearranged in
the following form
\begin{equation}
 V_{\mathcal{O(V}^{-3-p})}=(C_1+C_2\tau_3)\cos(2\pi\rho_3)e^{-2\pi\tau_3}\, ,
\end{equation}
where $C_1$ and $C_2$ can be identified from the original expression of
$V_{\mathcal{O(V}^{-3-p})}$ easily. It is obvious that the minimal of
$V_{\mathcal{O(V}^{-3-p})}$ lies at
\begin{equation}
 \langle\tau_3\rangle=\frac{1}{2\pi}-\frac{C_1}{C_2}
\end{equation}
and $\rho_3 \in \mathbb{Z}$ or $\rho_3 \in \mathbb{Z}/2$ depends on the specific
value of $C_1$ and $C_2$. Unfortunately, under the relation of (\ref{eq:id1}),
we can show that $C_1=0$, namely $\langle\tau_3\rangle=\frac{1}{2\pi}$, which is
manifest that $\tau_3$ is out of the K$\ddot{\textrm{a}}$hler cone, since it
will lead to the negative value of the volume of W-surface.

After checking the procedure of the stabilisation, we find, if we use the
precise
relation derived from $\frac{\partial V}{\partial\tau_i}=0$ instead of the
approximation relation (\ref{eq:id1}), the value of $C_1$ is indeed non-zero.
However, it seems that $C_1$ and $C_2$ have mostly  the same sign. Even when
they are in different sign, with $C_1 \ll C_2$,  it can not solve the
problem.

In order to solve the problem, we need introduce more parameters by using the
racetrack superpotential as suggested in \cite{1105.2107}. In this case, the
superpotential reads as
\begin{eqnarray}
 W=W_0&+&A_1 e^{-a_1 T_1}+A_1 A_6 e^{-a_1 T_1-2 \pi(T_3-T_1-T_2)}
 +A_2 e^{-a_2 T_2}+A_2 A_6 e^{-a_2 T_2-2 \pi(T_3-T_1-T_2)} \nonumber \\
&-&B_1 e^{-b_1 T_1}-B_1 B_6 e^{-b_1T_1-2\pi(T_3-T_1-T_2)}
-B_2 e^{-b_2 T_2}-B_2 B_6 e^{-b_2 T_2-2 \pi(T_3-T_1-T_2)}\, ,\nonumber \\
\label{eq:race}
\end{eqnarray}
where $a_i=\frac{2\pi}{m_i}, b_i=\frac{2\pi}{l_i}, m_i, l_i \in  \mathbb{Z}_{+},
i=1,2$. Taking this racetrack superpotential, we can repeat the previous
procedures. First of all, the possible orders in $\mathcal{V}$ in the scalar
potential Eq.(\ref{eq:Pot}) are the same as before, namely (\ref{eq:ord}), and
also only the terms whose order involves $p$ are relevant to the modulus
$\tau_3$.

The leading order in $\mathcal{V}$ of the scalar potential reads
\begin{eqnarray}
 V_{\mathcal{O(V}^{-3})}=&&\sum_{i=1}^2
\frac{1}{\mathcal{V}}\left[\frac{8}{3\beta_i}
 \sqrt{\tau_i} \left(a_i^2 A_i^2 e^{-2 a_i \tau_i}-2 a_iA_i b_i B_i
 e^{-a_i\tau_i-b_i\tau_i}\cos\left(a_i \rho_i-b_i \rho_i\right)+b_i^2 B_i^2 
 e^{-2b_i\tau_i}\right)\right]   \nonumber\\
   &&+\sum_{i=1}^2 \frac{1}{\mathcal{V}^2}\left[4W_0 \tau_i 
   \left(b_i B_i e^{-b_i \tau_i} \cos \left(b_i
   \rho_i\right)-a_i A_i e^{-a_i \tau_i} \cos \left(a_i
   \rho_i\right)\right)\right]+\frac{3W_0^2}{4\mathcal{V}^3}\hat{\xi}\, .
\end{eqnarray}
As before, let us find out the stabilised value of the axions $\rho_i,i=1,2$,
where the
relevant derivatives are:
\begin{eqnarray}
 \frac{\partial V}{\partial
\rho_i}&=&\frac{1}{\mathcal{V}}\left[\frac{16}{3\beta_i} a_i
 A_i b_i B_i \sqrt{\tau_i}(a_i-b_i) e^{-a_i \tau_i-b_i \tau_i}
   \sin\left(a_i \rho_i-b_i \rho_i\right)\right]\nonumber \\
   &&+\frac{1}{\mathcal{V}^2}\left[4 W_0 \tau_i 
   \left(b_i^2 B_i e^{-b_i \tau_i} \sin \left(b_i\rho_i\right)-a_i^2 A_i 
   e^{-a_i\tau_i} \sin\left(a_i \rho_i\right)\right)\right]\, ,\nonumber \\
 \frac{\partial^2 V}{\partial
\rho_i^2}&=&\frac{1}{\mathcal{V}}\left[\frac{16}{3\beta_i} a_i A_i b_i B_i
\sqrt{\tau_i} (a_i-b_i)^2 e^{-a_i \tau_i-b_i \tau_i} 
 \cos \left(a_i \rho_i-b_i \rho_i\right)\right]\nonumber \\
   &&+\frac{1}{\mathcal{V}^2}\left[4 W_0 \tau_i 
   \left(b_i^3 B_i e^{-b_i \tau_i} \cos \left(b_i \rho_i\right)-a_i^3 A_i 
   e^{-a_i\tau_i} \cos\left(a_i \rho_i\right)\right)\right]\, .
\end{eqnarray}
Note that $\frac{\partial V}{\partial \rho_i}$ vanishes at $\rho_i=0$, which is 
a minimum, if
\begin{eqnarray}
 \frac{\partial^2 V}{\partial\rho_i^2}\mid_{\rho_i=0}&=&\frac{1}{\mathcal{V}}
 \left[\frac{16}{3\beta_i}a_iA_ib_iB_i\sqrt{\tau_i}(a_i-b_i)^2
e^{-a_i\tau_i-b_i\tau_i} 
 \right]\nonumber \\ &+&\frac{1}{\mathcal{V}^2}\left[4 W_0 \tau_i 
   \left(b_i^3 B_i e^{-b_i \tau_i}-a_i^3 A_i e^{-a_i\tau_i} \right)\right]>0.
   \label{eq:con}
\end{eqnarray}
For simplicity, we assume this condition to be true. Now we start analysing
the stabilisation of $\tau_i, i=1,2$. As before, we first take
the limit of $a_i\tau_i\gg1$, then the vanishing of $\frac{\partial V}{\partial
\tau_i}$ implies
\begin{equation}
 e^{-b_i\tau_i}=\frac{3\beta_i W_0\tau_i^{1/2}}{4Z_i\mathcal{V}},
\quad\textrm{with} \quad Z_i=b_iB_i-A_i a_i e^{-n_i\tau_i}\, ,
 \label{eq:id3}
\end{equation}
where we have written $a_i=b_i+n_i$. In addition $Z_i>0$ can ensure that the
condition (\ref{eq:con}) is satisfied. Inserting this identity into the equation
$\frac{\partial V}{\partial\mathcal{V}}=0$, this equation reduces to
\begin{equation}
 2\left(\beta_1\tau_1^{3/2}+\beta_2\tau_2^{3/2}\right)=\hat{\xi},
\end{equation}
which is the same as the case with minimal superpotential. Therefore we obtain
the same value for $\langle\tau_i\rangle$ as before, and the value of the volume
$\mathcal{V}$ can be determined by the identity (\ref{eq:id3}).

The next step is to fix the value of $\tau_3$ and $\rho_3$ by the scalar
potential $V_{\mathcal{O(V}^{-3-p})}$, which read as
{\footnotesize
\begin{eqnarray}
 &&V_{\mathcal{O(V}^{-3-p})}\nonumber\\
 =&&\bigg\{\frac{16}{3\beta_1\mathcal{V}}\tau_1^{1/2}\cos(2\pi\rho_3)
 e^{-2\pi\tau_3}\left[a_1^2 A_1^2 A_6 e^{2(\pi-a_1)\tau_1+2\pi\tau_2}
 -2\pi a_1 A_1^2 A_6 e^{2(\pi -a_1) \tau_1+2\pi\tau_2}\right. \nonumber\\
 &-&2\pi a_1 A_1 A_2 A_6 e^{-(a_1-2 \pi ) \tau _1-(a_2-2 \pi ) \tau _2}
   -a_1A_1A_6b_1B_1e^{2\pi\tau_2-\tau_1(a_1+b_1-2\pi)}+2\pi A_1 A_6 b_1B_1
   e^{2\pi \tau_2-\tau_1(a_1+b_1-2\pi)}\nonumber\\
  &-&a_1A_1b_1B_1B_6 e^{2\pi\tau_2-\tau_1(a_1+b_1-2\pi)}+2\pi a_1 A_1 B_1 B_6
e^{2\pi\tau_2-\tau_1(a_1+b_1-2\pi)}+2\pi a_1A_1B_2B_6
   e^{-(a_1-2\pi)\tau_1-(b_2-2\pi)\tau_2}\nonumber\\
   &+&2\pi A_2A_6b_1B_1 e^{-(a_2-2\pi)\tau_2-(b_1-2\pi)\tau_1}+b_1^2 B_1^2
B_6 e^{2(\pi-b_1)\tau_1+2\pi\tau_2}-2\pi b_1B_1^2 B_6 
   e^{2(\pi-b_1)\tau_1+2\pi \tau_2}\nonumber\\
   &-&\left.2\pi b_1 B_1 B_2 B_6
   e^{-(b_1-2 \pi ) \tau _1-(b_2-2 \pi ) \tau _2}\right]\nonumber\\
   &+&\frac{1}{\mathcal{V}^2} 4 W_0 \tau_1 e^{-2\pi\tau_3}\cos(2\pi\rho_3)
   \left[a_1 A_1 A_6 e^{2\pi \tau_2-(a_1-2 \pi)\tau_1}-2\pi A_1 A_6 
   e^{2\pi\tau_2-(a_1-2\pi)\tau_1}-2\pi A_2 A_6
e^{2\pi(\tau_1+\tau_2)-a_2\tau_2}
   \right.\nonumber\\
   &-&\left. b_1 B_1 B_6 e^{2\pi \tau_2-(b_1-2\pi)\tau_1}+2\pi B_1 B_6 
   e^{2\pi \tau_2-(b_1-2\pi)\tau_1}+2\pi B_2 B_6
e^{2\pi(\tau_1+\tau_2)-b_2\tau_2}\right]+(\textrm{subscripts:} 1\leftrightarrow
2)\bigg\}\nonumber\\
&+&\frac{1}{\mathcal{V}^2} 4 W_0 \pi\tau_3
e^{-2\pi\tau_3}\cos(2\pi\rho_3)\left[2 A_1 A_6 
e^{2\pi\tau_2-(a_1-2\pi)\tau_1}+2 A_2 A_6 e^{2\pi(\tau_1+\tau_2)-a_2\tau_2}
-2 B_1 B_6 e^{2\pi\tau_2-(b_1-2\pi)\tau_1}\right.\nonumber\\
&-&\left.2 B_2 B_6 e^{2\pi(\tau_1+\tau_2)-b_2\tau_2}\right],
\end{eqnarray}
}
where we have used the stabilised value of the axions $\rho_i$ in this
expression. The scalar potential in the order of $\mathcal{V}^{-3-p}$ can be
rearranged in the form,
\begin{equation}
 V_{\mathcal{O(V}^{-3-p})}=(C_1^\prime+C_2^\prime\tau_3)\cos(2\pi\rho_3)e^{
-2\pi\tau_3}
\end{equation}
as before and the constant $C_1^\prime, C_2^\prime$ can be easily identified
from the original expression of $V_{\mathcal{O(V}^{-3-p})}$. Hence its minimum
lies at $\langle\tau_3\rangle=\frac{1}{2\pi}-\frac{C_1^\prime}{C_2^\prime}$, and
$\rho_3 \in \mathbb{Z}$ or $\rho_3 \in \mathbb{Z}/2$ depends on the specific
value of $C_1^\prime$ and $C_2^\prime$. Unfortunately, when we take the
approximate relation (\ref{eq:id3}), $C_1^\prime$ will vanish again. However in
this racetrack superpotential case, we indeed can get  a suitable ratio of
$C_1^\prime$ and $C_2^\prime$ by solving the equation $\frac{\partial
V_{\mathcal{O(V}^{-3})}}{\partial\tau_i}=0$ exactly instead of the approximate
relation (\ref{eq:id3}). In this case the relation will be modified as
\begin{equation}
 e^{-b_i\tau_i}=\frac{3\beta_i W_0 \tau_i^{1/2}}{4Z_i
\mathcal{V}}f_i^{\textrm{corr}}
 \label{eq:id4}
\end{equation}
where
\begin{equation}
 f_i^{\textrm{coor}}=1-\frac{3\epsilon_i}{-\epsilon_i+1+n_i\left(\frac{1}{b_i}
-\frac{B_i}{Z_i}\right)} \quad \textrm{with}\quad
\epsilon_i:=\frac{1}{4b_i\tau_i}\ll1\, .
\end{equation}
One subtle thing is that we can not solve the fixed values of $\tau_i,i=1,2$
analytically any more, so we have to solve them numerically.  Then we obtain the
stabilised value of the volume $\mathcal{V}$ by using the identity
(\ref{eq:id4}).
Note that although the relation has been modified, the stabilised value of the
volume $\mathcal{V}$ and $\tau_1, \tau_2$ are still independent of $\tau_3$.
After that, we can get the stabilised value of $\tau_3$ numerically by
minimising the scalar potential $V_{\mathcal{O(V}^{-3-p})}$.  We can also
determine the value of $p$ as follows:
\begin{equation}
 p=\frac{l_i(\tau_3-\tau_1-\tau_2)}{\tau_i},\quad\forall i=1,2\, .
\end{equation}
Here we must stress that the definition of the $p$ is different from the one in
\cite{1105.2107}, as we have a much more complex structure of
$V_{\mathcal{O(V}^{-3-p})}$. This definition originates from the direct
comparison
of these two relations, $e^{b_i\tau_i}\propto\mathcal{V},
e^{2\pi(\tau_3-\tau_1-\tau_2)}\propto\mathcal{V}^p$. 

We will give some benchmark models in table \ref{tab:ben} and the fixed values
of divisor volume moduli corresponding to the respective benchmark models in
table \ref{tab:res}, to show that this procedure of moduli stabilisation does
work consistently.

\begin{table}[h]
  \noindent\makebox[\textwidth][c]{
   \begin{tabular}{c||cccccccc||ccc}
Nos.&$g_s$&$a_i$&$b_i$&$A_i$&$B_i$&$A_6$&$B_6$&$W_0$&$f_i^{\textrm{corr}}$&
$Z_i$&$p$\\
   \hline
  I&0.1&$\frac{2\pi}{7}$&$\frac{2\pi}{8}$&0.1&0.1&7&4.5&1&0.57&0.021&1.00 \\
  II&0.1&$\frac{2\pi}{8}$&$\frac{2\pi}{7}$&0.1&0.2&6&4.37&1&0.79&0.065&0.51\\
   \hline
  III&0.01&$\frac{2\pi}{8}$&$\frac{2\pi}{7}$&0.1&8&21&7.96&1&0.97&4.80&0.21\\
  IV&0.01&$\frac{2\pi}{7}$&$\frac{2\pi}{8}$&7&0.5&12&5.45&1&0.96&0.19&0.57
   \end{tabular}
 }
 \caption{\label{tab:ben}Parameters for four benchmark models}
\end{table}

\begin{table}[h]
  \noindent\makebox[\textwidth][c]{
   \begin{tabular}{c||cccc||c}
Nos.&$\tau_i$&$\tau_3$&$\rho_3$&$\mathcal{V}$&$M_s$(GeV)\\
   \hline
  I&4.07&8.65&$\mathbb{Z}$&473.33&$1.30\times10^{16}$ \\
  II&3.44&7.13&$\mathbb{Z}/2$&174.76&$2.13\times10^{16}$ \\
   \hline
  III&31.02&62.96&$\mathbb{Z}/2$&$4.92\times10^{11}$&$4.02\times10^{11}$ \\
  IV&31.17&64.56&$\mathbb{Z}$&$4.27\times10^{11}$&$4.32\times10^{11}$
   \end{tabular}
 }
 \caption{\label{tab:res}The fixed values of divisor volume moduli corresponding
to the respective benchmark models}
\end{table}

The string scale in the table is defined as usual
$M_s=\frac{M_p}{\sqrt{4\pi\mathcal{V}}}$. And we have set the constants $\alpha,
\beta_1, \beta_2$ in the expression of volume as
$\alpha=\frac{1}{6},\beta_1=\beta_2=\frac{\sqrt{2}}{3}$, so that the volume is
just the same as the explicit example presented in section 3.1. 

Observing the table of benchmark models, one should note that we always choose
the symmetric value of $\tau_1$ and $\tau_2$. If we only consider the
stabilisation of the K$\ddot{\textrm{a}}$hler moduli by minimising the scalar
potential, it seems that this kind of symmetry is not necessary. However when we
consider the geometrical and topological properties of these two divisors in the
Calabi-Yau threefold, it seems that this kind of symmetry indeed exists. Of
course, one may search the possible parameters, which breaks this kind of
symmetry.

\textbf{String loop corrections:} Except the $\alpha^\prime$-corrections
and the non-perturbative corrections mentioned above, there are also corrections
to
the scalar potential from the string loop effects. In this final part,
let us  analyse the affect of this string loop correction to our procedure of
moduli stabilisation.

The string loop corrections to the scalar potential read as \cite{0708.1873}
\begin{equation}
 \delta V_{(g_s)}=\sum_i \left(g_s^2 \left(\mathcal{C}_i^{KK}\right)^2
K_{ii}-2\delta K_{(g_s),\tau_i}^W\right)\frac{W_0^2}{\mathcal{V}^2}
\quad\textrm{with}\quad \delta
K_{(g_s),\tau_i}^W=\sum_{l}\frac{\mathcal{C}_i^W}{(a_{il}t^l)\mathcal{V}}\,  ,
\end{equation}
where $\mathcal{C}_i^{KK}$ and $\mathcal{C}_i^W$ are constants which depend on
the complex structure moduli, $K_{ii}$ is the K$\ddot{\textrm{a}}$hler metric
for $\tau_i$ and $a_{il}t^l$ denotes a linear combination of the basis 2-cycle
volumes $t^l$. The first part of the correction comes from the exchange of
closed strings which carry Kaluza-Klein momentum between D7 and D3-branes and
the second part comes from the exchange of winding strings between intersecting
stacks of D7-branes. Applying this to the explicit example presented in section
3.1, since there are no intersection between every pair of the divisors $D_2,
D_4, D_5$, which are wrapped by the D7-branes, it is  obvious that $\delta
K_{(g_s),\tau_i}^W=0$. Therefore, in our constructions, the string loop
corrections to the scalar potential are:
\begin{equation}
 \delta V_{(g_s)}=g_s^2
\frac{W_0^2}{\mathcal{V}^2}\left[\frac{3\beta_1\left(\mathcal{C}_1^{KK}\right)^2
}{8\mathcal{V}\sqrt{\tau_1}}+\frac{3\beta_2\left(\mathcal{C}_2^{KK}\right)^2}{
8\mathcal{V}\sqrt{\tau_2}}+\frac{\left(\mathcal{C}_3^{KK}\right)^2}{4\tau_3^2}
\right]\, .
\label{eq:loop}
\end{equation}
Since $g_s<1$, the LARGE volume scenario will be safe as long as
$W_0\sim\mathcal{O}(1)$. One can also see the effects of the string loop
corrections to
various moduli stabilisation mechanism in the appendix of \cite{0708.1873}. For
the stabilisation of $\tau_3$, the leading order scalar potential relevant to
the divisor $\tau_3$ seems from this string loop corrections, since
superficially it scales as $\mathcal{V}^{-2}$, but we can estimate the scale of 
the last term in Eq.(\ref{eq:loop}), denoted as $\delta V_{(g_s)}^{(3)}$, by 
using the 1-loop Coleman-Weinberg potential as in
\cite{1105.2107}, which is showed that $\delta V_{(g_s)}^{(3)}\sim \Lambda^2
STr(M^2) \sim (M_{KK}^{6D})^2 m_{3/2}^2 \sim \frac{\tau_3}{\mathcal{V}^4}$,
where the cut-off $\Lambda$ in the 1-loop Coleman-Weinberg potential given by
the 6D Kaluza-Klein scale $M_{KK}^{6D} \sim
M_P\frac{\sqrt{\tau_3}}{\mathcal{V}}$ and $m_{3/2} \sim
\frac{M_P}{\mathcal{V}}$. Considering that the leading order term relevant to 
$\tau_3$ in the poly-instanton potential scales as $\mathcal{V}^{-3-p}$, we can 
conclude that the string loop corrections of the scalar potential will not spoil
 the procedure of the moduli stabilisation.

\section{Conclusions}
In this paper, we have presented one consistent procedure to generate the
superpotential in types of Eq.(\ref{eq:W}), including the gaugino condensation
and poly-instanton effects, in type IIB orientifold compactification. And then
we use this kind of superpotential as well as the $\alpha^\prime$-corrections to
the K\"ahler potential to stabilise all four K\"ahler moduli, where the volume
form of the compact Calabi-Yau is in the general form Eq.(\ref{eq:GK}). For this
purpose we first searched all the possible Calabi-Yau threefolds which have one
W-surface to support  the poly-instanton effects from the 158 examples of
reflexive lattice polytopes, which admit a K3-fibred Calabi-Yau hypersurface
in \cite{1107.0383}. We find that only 23 of them admit a W-surface, where
the result has been listed in the appendix. Then we analysed all the topological
and geometrical conditions of the non-perturbative superpotential induced by the
gaugino condensation and poly-instanton effect for one explicit Calabi-Yau
threefold. Finally we systematically studied the stabilisation procedure  and
discussed that this procedure is safe from the string-loop corrections.

One advantage of K3-fibred Calabi-Yau threefolds in type IIB theory is that
the extra dimensions can be anisotropic. Thus we can try to embed the
supersymmtric extra dimensional models of particle physics into this frame, such
as \cite{1105.2107}. However, one of the constraints for this kind of embedding
enforces the string scale to be around TeV-scale, namely the volume of the
compactified manifold should be around $10^{28}$. In our procedure of moduli
stabilisation, the volume is proportion to the exponential of $\tau_i$, whose
value is mainly determined by the value of $\hat{\xi}$. As the Calabi-Yau
threefold for compactification has been uniquely chosen, the only possible
change is in the string coupling $g_s$. If we demand the volume to be around
$10^{28}$, the string coupling should be around $g_s \sim 0.004$, which is
unnaturally small. Of course for the embedding of a concrete particle
physics model, not all the K$\ddot{\textrm{a}}$hler moduli should be fixed by
this procedure, because some of moduli or their linear combinations can be fixed
by the $D$-terms, which are generated from the D7-brane gauge theory in the
visible sectors.

In addition, we can also expect that this procedure can be used in the string
cosmology, especially for various inflationary models, in which the
K$\ddot{\textrm{a}}$hler moduli will serve as  inflaton field in the single
field inflation scenarios, or both inflaton and curvaton(or light modulating
field) in the double-field inflation scenarios (for an overview of this point
see \cite{1108.2659}). To constrain the K3-fibred Calabi-Yau threefold in the
single field inflation scenarios, either one of the del Pezzo surfaces or the
K3-divisor can serve as the inflaton, as in \cite{0808.0691,1110.6182}. For the
double-field inflation scenarios, more precisely for the curvaton mechanism, one
of the del-Pezzo surfaces serves as the inflaton and the fibre K3 divisor
servers as curvaton \cite{1005.4840}, and for the modulation mechanism, the
fibre K3 divisor can serve as the inflaton, while the W-surface can serve as the
light modulating field \cite{1202.4580}.  The explicit example present in this
paper can either be used for the single-field inflation or for the double-field
scenario, since at the leading order, the structure of the scalar potential is
similar as the one in these papers. For the same reason, one can also expect
that the explicit example present in this paper to be used in some quintessence
models for dark energy, in which the quintessence field can be identified as the
fibre K3 divisor \cite{1203.6655}.

Finally we need to point out that the racetrack superpotential used in the
procedure of moduli stabilisation in this paper is just an assumption, instead
of a concrete construction as for the superpotential Eq.(\ref{eq:W}). To get a
racetrack superpotential, we need introduce the gauge flux on the D7-branes to
split the original gauge group into two parts with different ranks, and then
have gaugino condensation on both of them. Note that the different ranks is not
a general requirement for generating racetrack superpotenial, but it's necessary
to moduli stabilisation, since we need that $a_i \neq b_i$. However turning on
the gauge flux will destroy the favourable zero-modes of instantons, so we must
carefully choose the gauge flux to cancel the extra zero-modes. That is much
more complicated, and beyond the simple setups in this paper, so we leave it to
the later works.

\acknowledgments{We thank P. Shukla and X. Gao for useful discussions on the
non-perturbative effects to the superpotential. And also we thank T. Weigand and
M. Cicoli for very useful comments on the manuscript. X. Zhang is supported by
the MPG-CAS Joint Doctoral Promotion Program.}

\newpage
\appendix
\section{List of all the K3 fibrations with del Pezzo and W-surface}
In this appendix we give all the 23 four dimensional reflexive lattice
polytopes, which admit a K3-fibred Calabi-Yau hypersurface with four
K$\ddot{\textrm{a}}$hler moduli in which at least one del Pezzo surface and one
W-surface.

We must stress that the difference between the weight matrices of the polytopes
listed here and those listing in \cite{1107.0383}, is due to the fact that the
reflexive lattice polytopes which can be defined without having to specify all
the weights. But in order to make comparisons between each other, we will
preserve the mark number.

\begin{table}[h]
\begin{center}
\noindent\makebox[\textwidth][c]{\small
\begin{tabular}{c||c|c|c|c|c|c|c|c}
 Nos.&
$\sum_A$&$\omega_i^A$&$\sum_B$&$\omega_i^B$&$\sum_C$&$\omega_i^C$&$\sum_D$&
$\omega_i^D$\\
 \hline \hline
 1& 8&2 1 2 1 2 0 0 0& 8&2 1 1 0 2 0 0 2& 8&2 0 1 1 2 0 2 0& 4&1 0 1 0 1 1 0 0\\
 3&12&2 1 6 1 2 0 0 0&12&2 1 6 0 1 0 2 0&12&2 0 6 1 1 0 0 2& 6&1 0 3 0 1 1 0 0\\
 6& 8&1 2 2 0 2 1 0 0& 5&1 1 1 0 1 0 1 0& 2&1 0 0 1 0 0 0 0& 4&0 1 1 0 1 0 0 1\\
10&12&4 2 1 2 3 0 0 0& 6&2 1 1 0 1 1 0 0& 6&2 0 1 0 1 0 2 0& 3&1 0 0 0 0 0 1 1\\
12&12&6 2 0 1 2 1 0 0& 8&4 1 0 1 1 0 1 0& 6&3 1 0 0 1 0 0 1& 4&2 0 1 1 0 0 0 0\\
17&14&7 2 0 3 1 1 0 0&16&8 0 1 4 2 1 0 0& 8&4 0 0 2 1 0 0 1& 4&2 0 0 1 0 0 1 0\\
21&24&8 1 12 0 2 1 0 0&18&6 0 9 1 1 1 0 0&12&4 0 6 0 1 0 1 0&6&2 0 3 0 0 0 0 1\\
32& 6&3 1 1 0 1 0 0 0&10&5 0 1 1 2 0 1 0& 8&4 0 0 1 0 1 2 0& 4&2 0 0 0 0 0 1 1\\
33& 8&1 0 1 2 2 0 2 0& 5&1 0 0 1 1 1 1 0& 3&0 1 0 0 1 0 1 0& 4&0 0 0 1 1 0 1 1\\
34& 6&2 1 0 1 0 1 1 0& 6&2 1 0 0 1 2 0 0& 3&1 0 1 1 0 0 0 0& 3&1 0 0 0 0 1 0 1\\
35& 6&2 1 1 1 0 1 0 0& 6&2 1 2 0 0 0 1 0& 6&1 1 0 1 1 2 0 0& 3&0 0 0 0 1 1 0 1\\
41& 8&4 1 0 0 0 1 2 0& 4&2 0 1 0 1 0 0 0&10&5 0 0 2 1 1 1 0& 4&2 0 0 0 0 0 1 1\\
56& 3&1 0 0 0 0 1 1 0& 8&0 1 1 2 0 2 2 0& 7&0 1 0 1 1 2 2 0& 4&0 0 0 1 0 1 1 1\\
60& 8&4 1 0 0 1 1 1 0&10&5 0 1 0 1 1 2 0& 8&4 0 0 1 2 1 0 0& 4&2 0 0 0 1 0 0 1\\
62&10&5 1 1 0 1 0 2 0&12&6 2 0 0 1 1 2 0& 6&3 1 0 0 0 0 1 1& 8&4 0 1 1 1 0 1 0\\
63&10&5 1 1 0 2 0 1 0& 8&4 1 0 0 0 1 2 0& 4&2 0 0 1 1 0 0 0& 4&2 0 0 0 0 0 1 1\\
69& 8&4 1 2 0 1 0 0 0& 6&3 1 1 0 0 1 0 0& 8&4 0 1 1 1 0 1 0& 4&2 0 1 0 0 0 0 1\\
71& 6&3 1 0 0 0 1 1 0& 8&4 0 1 0 1 1 1 0& 6&2 0 0 1 1 1 1 0& 7&3 0 0 0 1 1 1 1\\
77& 6&3 0 1 1 0 0 1 0& 4&2 0 0 0 1 1 0 0& 3&1 0 0 0 0 1 0 1& 2&0 1 0 0 0 1 0 0\\
79&12&6 1 0 0 2 1 2 0& 4&2 0 1 0 0 0 1 0& 6&3 0 0 1 1 1 0 0& 6&3 0 0 0 1 0 1 1\\
95&12&6 1 0 0 2 2 1 0& 8&4 0 1 0 1 1 1 0& 4&2 0 0 1 0 1 0 0& 6&3 0 0 0 1 1 0 1\\
99&12&6 1 2 0 1 0 2 0&10&5 1 1 0 0 1 2 0& 6&3 0 1 0 0 0 1 1& 4&2 0 0 1 0 0 1 0\\
106&8&1 1 4 0 1 0 1 0& 4&1 0 2 0 0 1 0 0& 8&0 1 4 1 0 0 2 0& 4&0 0 2 0 0 0 1 1
\end{tabular}
}
\end{center}
\end{table}

\bibliographystyle{utphys}


\providecommand{\href}[2]{#2}\begingroup\raggedright\endgroup

\end{document}